\documentclass{ws-procs975x65}

\newcommand{\be}{\begin{equation}}
\newcommand{\ee}{\end{equation}}
\newcommand{\bea}{\begin{eqnarray*}}
\newcommand{\eea}{\end{eqnarray*}}
\newcommand{\bdm}{\begin{displaymath}}
\newcommand{\edm}{\end{displaymath}}
\newcommand{\D}{{\rm d}}
\newcommand{\E}{{\rm e}}
\newcommand{\I}{{\rm i}}

\begin{document}

\title{Quantum black hole without singularity}

\author{Claus Kiefer} 

\address{Institute for Theoretical Physics, University of Cologne,\\
Z\"ulpicher Strasse 77, 50937 K\"oln, Germany\\
E-mail: kiefer@thp.uni-koeln.de}

\begin{abstract}
We discuss the quantization of a spherical dust shell in a rigorous
manner. Classically, the shell can collapse to form a black hole with a
singularity. In the quantum theory, we 
construct a well-defined self-adjoint extension
for the Hamilton operator. As a result, the evolution is unitary and
the singularity is avoided. If
we represent the shell initially by a narrow wave packet, it will
first contract until it reaches the region where classically a black
hole would form, but then re-expands to infinity. In a way, the state can be
interpreted as a superposition of a black hole with a white hole. 
\end{abstract}

\keywords{Black holes; canonical quantum gravity; singularity
  avoidance}

\bodymatter


\vskip 1cm

Among the problems to be addressed by any sensible quantum theory of
gravity is the issue of black hole evaporation. Hawking's
semiclassical calculations break down when the black hole approaches
the Planck regime, so the question arises whether there will be only
thermal radiation left behind or not. In other words, the issue is
whether there is unitarity violation for the full system or not;
loosely speaking, whether ``information'' is lost or not.  
A related question is the quantum fate of the singularity predicted by general
relativity. 

Since there is not yet agreement about the correct theory of quantum
gravity, an analysis must be based on one of the existing approaches.
\cite{OUP} Typically, one can either perform a formal and limited
analysis of the full equations or perform a rigorous analysis of a
simple model. Here, we choose the second way. We briefly review the
exact quantization of a thin shell consisting of matter with vanishing
rest mass. Classically, such a ``null dust shell'' can collapse to
form a black hole or expand from a white hole. As we shall see, there
is no black or white hole in the quantum version, but only a
superposition (in some sense) of them, which avoids the formation of a
singularity. This model was discussed already some time ago 
[\refcite{HK01a,Ha01,HK01b,AH05,HaReview}], and we refer the reader to these
references for technical details. Our work is presented here
because of some recent interest in this direction, see [\refcite{HR15,Ba14}]
as well as other contributions to the BH6 session and the references
therein.  

Our approach is canonical quantum gravity (quantum geometrodynamics). 
\cite{OUP} We use the method of reduced quantization and separate the
variables and their canonical momenta into pure gauge degrees of
freedom (``embedding variables'') and physical degrees of freedom. The
general existence of this ``Kucha\v{r} decomposition'' can be shown by
making a transformation to the standard ADM phase space of general
relativity. \cite{HK00} In this construction, the presence of a formal
``background manifold'' plays a crucial role.\cite{HK01a}

Let us first present the classical model.
 All physically distinct solutions can be labelled 
by three parameters: a parameter $\eta\in
\{-1,+1\}$ that distinguishes between the outgoing ($\eta = +1$) and
the ingoing
($\eta = -1$) null surfaces; the asymptotic time of the surface,
that is, the 
retarded time $u=T-R\in (-\infty,\infty)$ for $\eta = +1$, and the advanced
time $v=T+R\in (-\infty,\infty)$ for $\eta = -1$; and the mass $M\in
(0,\infty)$. An ingoing shell creates a black-hole (event)
horizon at $R=2M$ 
and ends up in the singularity at $R=0$. The outgoing shell starts from the
singularity at $R=0$ and emerges from a white-hole (particle) horizon at
$R=2M$. The metric reads
\bdm
  \D s^2  =  -A(U,V)\,\D U\,\D V + R^2(U,V)(\D\theta^2 +
  \sin^2\theta \,\D\phi^2).
\edm
 From the demands that this metric be continuous at the position of the
 shell and flat under the shell,
the coefficients $A$ and $R$
are uniquely defined for any physical situation defined
by the variables $M$ (the energy of the shell), $\eta$,
and $w$ (the location of the shell, where $w=u$ for the outgoing
and $w=v$ for the ingoing case).\cite{HK01a}

In the standard (ADM) formulation, a general spherically symmetric
line element assumes the form
\bdm
  \D s^2 = -N^2\,\D\tau^2 + L^2(\D\rho + N^{\rho}\,\D\tau)^2 + 
R^2\,\D\Omega^2,
\edm
where $N$ is the lapse and $N^{\rho}$ the shift function. 
The shell is located at $\rho = {\mathbf r}$. The action reads 
\bdm
  S_0 = \int \D\tau\left[{\mathbf p}\dot{\mathbf r} + \int
  \D\rho\, (P_L\dot{L} + P_R\dot{R} - H_0)\right],
\edm
where the Hamiltonian is explicitly given by
\[
  H_0 = N{\mathcal H}_{\perp} + N^{\rho}{\mathcal H}_\rho\ + N_\infty E_\infty.
\] 
The Hamiltonian consists of the Hamiltonian constraint and the
diffeomorphism constraint,
\begin{eqnarray*}
 \!\!{\mathcal H}_{\perp} & = & \frac{L P_L^2}{2R^2} -
 \frac{P_L P_R}{R}  + 
 \frac{RR''}{L} - \frac{RR'L'}{L^2} 
 + \frac{R^{\prime 2}}{2L} - \frac{L}{2} + \frac{\eta{\mathbf
 p}}{L}\delta(\rho - {\mathbf r})\approx 0, 
 \\
\!\!  {\mathcal H}_\rho & = & P_RR' - P_L'L - {\mathbf
 p}\delta(\rho - {\mathbf r})\approx 0.  
\end{eqnarray*}
Such equations are discussed in 
[\refcite{Lo98}]; see also [\refcite{OUP}]. 

The explicit transformation to embedding variables (``Kucha\v{r}
decomposition'') then proceeds in two steps. First, in 
the transformation of the canonical coordinates $\mathbf r$, $\mathbf p$,
$L$, $P_L$, $R$, and $P_R$ {\em on the constraint surface}; and
second, in the {\em extension} of the functions $u$, $v$, $p_u$, $p_v$,
$U(\rho)$, $P_U(\rho)$, $V(\rho)$,
 and $P_V(\rho)$ {\em off the constraint surface}. This leads to the
 action 
\be
\label{action}
  S = \int \D\tau\left(p_u\dot{u} + p_v\dot{v} - np_up_v\right)
  + \int \D\tau\int_0^\infty \D\rho\, (P_U\dot{U} + P_V\dot{V} - H),
\ee
where $H = N^UP_U + N^VP_V$, and $n$, $N^U(\rho)$, and $N^V(\rho)$ are
Lagrange multipliers.
The first term in (\ref{action}) describes the physical part, and
the second term the embedding part.
 A crucial point is that the new phase space has non-trivial boundaries,
\bdm
  p_u \leq 0\ ,\quad p_v \leq 0,
\quad
 \frac{-u+v}{2} > 0.
\edm
The system has now been brought into a form which is suitable for
quantization.\cite{Ha01} We apply the method of {\em group
  quantization} to
\bdm
S_{\rm phys} = \int \D\tau\left(p_u\dot{u} + p_v\dot{v} - np_up_v\right).
\edm
In this approach, one has to make a
choice of a set of Dirac observables forming a Lie algebra.
This algebra generates a group of transformations which must respect all
boundaries, and it leads to self-adjoint operators for the observables; one
obtains, in particular, a {\em self-adjoint Hamiltonian} and thus a
{\em unitary dynamics}. 

A complete system of Dirac observables
is given by $p_u$, $p_v$, $D_u:= up_u$, and
$D_v:= vp_v$. The only non-vanishing Poisson brackets are
\bdm
\{ D_u,p_u\}=p_u\ , \quad \{ D_v,p_v\}=p_v.
\edm
 The {\em Hilbert space} is constructed from
complex functions $\psi_u(p)$ and $\psi_v(p)$, where
$p\in [0,\infty)$. The scalar product is defined by
\bdm 
  (\psi_u,\phi_u) := \int_0^\infty\ \frac{\D p}{p}\
    \psi^*_u(p)\phi_u(p) 
\edm
(and similarly for $\psi_v(p)$). It is useful to apply the
transformation
\begin{eqnarray*}
  t & = & (u+v)/2, \qquad r = (-u+v)/2,
\\
  p_t & = & p_u + p_v, \qquad \;\; p_r = -p_u + p_v.
\end{eqnarray*}
Upon quantization, one obtains the operator $-\hat{p}_t$,
which is self-adjoint and has a positive spectrum,
$-\hat{p}_t\varphi(p)=p\varphi(p)$, $p\geq 0$.  It is the
generator of time evolution (``Schr\"odinger evolution'')
and corresponds to the energy operator $E:= M$.

For the physical interpretation of the quantum theory, it is useful to
represent the thin shell by a wave packet and to study its
evolution. By construction, the evolution of the wave packet must be
unitary, so the wave packet cannot be ``lost''. We shall see that this
leads straightforwardly to singularity avoidance. 

It is convenient to start at $t=0$ with the wave packet
\bdm
  \psi_{\kappa\lambda}(p) :=
   \frac{(2\lambda)^{\kappa+1/2}}{\sqrt{(2\kappa)!}}
  p^{\kappa+1/2}\E^{-\lambda p},
\edm
where $\kappa$ and $\lambda$ are free parameters which describe the
size and the form of the wave packet. 
The expectation value for the energy and its variance with respect to
that state are
\bdm
\langle E\rangle_{\kappa\lambda}:=
\int_0^{\infty}\frac{\D p}{p}\ p\ \psi^2_{\kappa\lambda}(p)
=\frac{\kappa +1/2}{\lambda}
\edm
and
\bdm
\Delta E_{\kappa\lambda}=\frac{\sqrt{2\kappa +1}}{2\lambda},
\edm
respectively. 

Since the time evolution of the packet is generated by $-\hat{p}_t$,
one finds for $t>0$ in the $p$-representation immediately that
\bdm
  \psi_{\kappa\lambda}(t,p) = \psi_{\kappa\lambda}(p) \E^{-\I pt}.
\edm
Of more interest is the
exact time evolution in the $r$-representation,
for which one finds\cite{Ha01}
\be
  \Psi_{\kappa\lambda}(t,r) = \frac{1}{\sqrt{2\pi}}
  \frac{\kappa!(2\lambda)^{\kappa+1/2}}{\sqrt{(2\kappa)!}}
  \left[\frac{\I}{(\lambda +\I t +\I r)^{\kappa+1}}
   - \frac{\I}{(\lambda +\I t-\I r)^{\kappa+1}}\right]. 
\ee 
An important consequence of this result is that the wave function obeys
\be
\label{avoidance}
  \lim_{r\rightarrow 0}\Psi_{\kappa\lambda}(t,r)= 0.
\ee
The probability of finding the shell at
vanishing radius is thus zero! It is in this sense that
the {\em singularity is avoided
in the quantum theory}. The quantum shell bounces and re-expands, and
no event horizon forms. 
The expectation value and the variance of the shell radius
are given by
\bdm
\langle R_0\rangle_{\kappa\lambda}:= 2G\langle E\rangle_{\kappa\lambda}
=(2\kappa +1)\frac{l_{\rm P}^2}{\lambda}
\edm
and
\bdm
\Delta(R_0)_{\kappa\lambda}=2G\Delta E_{\kappa\lambda}=
\sqrt{2\kappa +1}\  \frac{l_{\rm P}^2}{\lambda},
\edm
respectively, where $l_{\rm P}$ is the Planck length and $G$ the
gravitational constant.  
Closer inspection shows that the wave packet can be squeezed
below its Schwarzschild radius if its energy is greater
than the Planck energy.\cite{Ha01} This is a genuine quantum effect!
In a sense, one can say that the initial quantum state develops into a
superposition of a black hole and a white hole state, although,
strictly speaking, neither a black nor a white hole develops; there
are no event horizons and no singularities. 

An initially collapsing shell will thus be re-expanding at a later
time. If the results of our very simple model were typical for the
generic situation of gravitational collapse, the collapsing object
would thus re-appear. But isn't this in conflict with observation?
After all, astrophysicists have never observed the time reversal of a
collapsing star. 

To answer this question, one must first calculate the time delay of
the re-expansion. Since the wave packet will collapse to a radius
equal or smaller than the Schwarzschild radius, the gravitational
redshift is expected to be very large. Such a calculation was performed in
[\refcite{AH05}], see also [\refcite{HR15}], but it is not yet clear
from the result whether there are conflicts with observation or not.
In order to capture the physics of a stellar collapse, probably a more
realistic model than the present one is needed. 

In our simple model, all local (field) degrees of freedom are
frozen. The model thus does not capture the effects of Hawking
radiation and black hole evaporation. To include those, one may
apply the methods of canonical quantization to a toy model such as the
one presented in [\refcite{KMM09}]. But so far, no rigorous treatment
comparable to the one here was achieved. If the feature of a
re-expanding object survived the inclusion of local degrees of
freedom, the information-loss problem for black holes would be solved
in a clear and intuitive manner: no information would be lost; it
would re-appear in the course of time. 

The vanishing of the wave function in the region of the classical
singularity can also be used as a criterion for the avoidance of
singularities in quantum cosmology, as proposed already by DeWitt in
1967.\cite{DeWitt} More recently, this criterion has been successfully
applied to a variety of singularities, including such exotic ones as
big brake or big freeze; see, for example, [\refcite{sing}], and 
references therein. 

We finally mention that a quantum cosmological
scenario of black and white hole superposition and singularity
avoidance also arises from a consistent implementation of a low-entropy
boundary condition for a small universe.\cite{KZ95} The ensuing arrow
of time there ensures that one can only observe an expanding universe
and objects collapsing to a black hole.

\section*{Acknowledgments}

I thank the organizers of session BH6 for inviting me to present
this topic in Rome and for providing a stimulating discussion atmosphere. 
I am also grateful to Petr H\'aj\'{\i}\v{c}ek for our collaboration on
this subject and for a critical reading of the manuscript.

\end{document}